%% file: letter.tex
\documentclass[aps,prl,preprintnumbers,reprint,superscriptaddress,longbibliography,amsmath,amssymb,twocolumn,10pt]{revtex4-1}
\usepackage{comment} 
\usepackage{graphicx}
\usepackage{xspace}
\usepackage{siunitx}
\usepackage[utf8]{inputenc}
\usepackage{dcolumn}
\usepackage{xcolor}
\definecolor{darkred}{rgb}{0.3,0,0}
\definecolor{darkblue}{rgb}{0,0,0.3}
\definecolor{firebrick}{rgb}{0.5,0.125,0.125}
\definecolor{darkgreen}{rgb}{0,0.3,0}
\usepackage[colorlinks=true,linkcolor=firebrick,citecolor=darkgreen,urlcolor=darkblue]{hyperref}

\newcommand{\sre}{s_{E}}

\newcommand{\rmu}{R_\mu}

\newcommand{\mrmu}{\langle \rmu \rangle}

\newcommand{\srmu}{s_{\mu}}

\newcommand{\lnrmu}{\ln\!\rmu}

\newcommand{\mlnrmu}{\langle \lnrmu \rangle}

\newcommand{\nmu}{N_\mu}

\newcommand{\xmax}{X_\text{max}}

\newcommand{\mxmax}{\langle \xmax \rangle}

\newcommand{\de}{\text{d}}

\def\sib{{Sibyll}\xspace}
\def\qgs{QGSJET\,II-04\xspace}

\def\epos{EPOS-LHC\,\xspace}

\newcommand{\dataStart}{January~1,~2004}
\newcommand{\dataEnd}{December~31,~2017}

\newcommand{\neventSel}{786}
\newcommand{\nevent}{281}

\newcommand{\enMuFrac}{14\%}

\newcommand{\minZenith}{62^\circ}
\newcommand{\maxZenith}{80^\circ}

\newcommand{\maxSystUncertainty}{$8$\%}

\newcommand{\mEnRes}{$(8.4 \pm 2.9)\,$\%\xspace}

\newcommand{\mRmuRes}{$(10 \pm 3)\,$\%\xspace}

\newcommand{\mesA}{1.86 \pm 0.02 \,(\mathrm{stat.}) ~_{-0.31}^{+ 0.36}\,(\mathrm{syst.})}

\newcommand{\mesB}{0.99 \pm 0.02 \,(\mathrm{stat.}) ~_{-0.03}^{ +0.03}\,(\mathrm{syst.})}

\newcommand{\sigFitB}{-0.10 \pm 0.04}
\newcommand{\sigFitA}{0.12 \pm 0.01}

\newcommand{\rescaleSib}{26\%}
\newcommand{\rescaleEpos}{35\%}
\newcommand{\rescaleQGS}{43\%}

\begin{document}

Published in Phys.\ Rev.\ Lett.\ as~\href{https://link.aps.org/doi/10.1103/PhysRevLett.126.152002}{{DOI:10.1103/PhysRevLett.126.152002}}

\title{Measurement of the fluctuations in the number of muons in extensive air showers with the Pierre Auger Observatory}

\date{\today}

\begin{abstract}
  We present the first measurement of the fluctuations in the number of muons in extensive air showers produced by ultrahigh energy cosmic rays. We find that the measured fluctuations are in good agreement with predictions from air shower simulations. This observation provides new insights into the origin of the previously reported deficit of muons in air shower simulations and constrains models of hadronic interactions at ultrahigh energies. Our measurement is compatible with the muon deficit originating from small deviations in the predictions from hadronic interaction models of particle production that accumulate as the showers develop.
\end{abstract}


\author{A.~Aab}
\affiliation{IMAPP, Radboud University Nijmegen, Nijmegen, The Netherlands}

\author{P.~Abreu}
\affiliation{Laborat\'orio de Instrumenta\c{c}\~ao e F\'\i{}sica Experimental de Part\'\i{}culas -- LIP and Instituto Superior T\'ecnico -- IST, Universidade de Lisboa -- UL, Lisboa, Portugal}

\author{M.~Aglietta}
\affiliation{Osservatorio Astrofisico di Torino (INAF), Torino, Italy}
\affiliation{INFN, Sezione di Torino, Torino, Italy}

\author{J.M.~Albury}
\affiliation{University of Adelaide, Adelaide, S.A., Australia}

\author{I.~Allekotte}
\affiliation{Centro At\'omico Bariloche and Instituto Balseiro (CNEA-UNCuyo-CONICET), San Carlos de Bariloche, Argentina}

\author{A.~Almela}
\affiliation{Instituto de Tecnolog\'\i{}as en Detecci\'on y Astropart\'\i{}culas (CNEA, CONICET, UNSAM), Buenos Aires, Argentina}
\affiliation{Universidad Tecnol\'ogica Nacional -- Facultad Regional Buenos Aires, Buenos Aires, Argentina}

\author{J.~Alvarez-Mu\~niz}
\affiliation{Instituto Galego de F\'\i{}sica de Altas Enerx\'\i{}as (IGFAE), Universidade de Santiago de Compostela, Santiago de Compostela, Spain}

\author{R.~Alves Batista}
\affiliation{IMAPP, Radboud University Nijmegen, Nijmegen, The Netherlands}

\author{G.A.~Anastasi}
\affiliation{Universit\`a Torino, Dipartimento di Fisica, Torino, Italy}
\affiliation{INFN, Sezione di Torino, Torino, Italy}

\author{L.~Anchordoqui}
\affiliation{Department of Physics and Astronomy, Lehman College, City University of New York, Bronx, NY, USA}

% ...

\collaboration{The Pierre Auger Collaboration}
\email{auger\_spokespersons@fnal.gov}
\homepage{http://www.auger.org}
\noaffiliation

\maketitle

\section{Introduction}
\label{sec:intro}

Ultrahigh energy cosmic rays (UHECRs) are particles coming from outer space, with energies exceeding $10^{18}\,$eV. They provide the only experimental opportunity to explore particle physics beyond energies reachable by Earth-based accelerators, which go up to cosmic ray energies of $9\times 10^{16}\,$eV.

The Pierre Auger Observatory~\cite{ThePierreAuger:2015rma} detects extensive air showers that are initiated by the UHECRs colliding with the nuclei in the atmosphere.
Information about UHECRs is extracted using simulations based on hadronic interaction models which rely on extrapolations of accelerator measurements to unexplored regions of phase space, most notably the forward and highest-energy region.
In addition, accelerator experiments at the highest energies either probe the interactions between protons
or of protons with heavy nuclei, while most interactions within air showers are between pions and light nuclei.

A further challenge is that the UHECR mass has to be measured despite
not being yet completely decoupled from the hadronic uncertainties.
The observable with the least dependence on hadronic interactions is the atmospheric depth at which the longitudinal development of the electromagnetic (EM) component of the shower reaches the maximum number of particles, namely $\xmax$~\cite{Linsley77a}.

In hadronic cascades the energy of each interacting particle is distributed among the secondaries, mostly pions.
Neutral pions rapidly decay into two photons feeding a practically decoupled electromagnetic cascade (other resonances decaying into $\pi^0$'s, electrons, and/or photons also contribute).
Charged pions (and other long-lived mesons like kaons) tend to further interact until their individual energies are below a critical value, below which they are more likely to decay.
Muons, which are products of hadronic decays, are thus predominantly produced in the final shower stages. 
In sufficiently inclined showers, the pure EM component is absorbed in the atmosphere and the particles that reach the ground (muons and muon decay products) directly sample the muon content~\cite{inclinedReco,Valino:2009dv}, reflecting the hadronic component of the shower.

Air showers are mainly detected at the Pierre Auger Observatory by the surface detector (SD), an array of water-Cherenkov detector stations, and the fluorescence detector (FD), consisting of 24 fluorescence telescopes.
By selecting the subsample of events reconstructed with both the SD and FD, and with zenith angles exceeding $62^\circ$, both the muon content and the energy of the shower are simultaneously measured.

The results obtained indicate that all the simulations underestimate the number of muons in the showers~\cite{Aab:2014pza,Sciutto:2019pqs}.
These analyses come with the caveat that they cannot distinguish a muon rescaling from a shift in the absolute energy scale of the FD measurement.
However, muon content and energy scale were disentangled in a complementary technique based on showers with zenith angles below $60^\circ$.
Using the longitudinal profile of the shower in the atmosphere obtained with the FD and the signals at the ground measured with the SD, it was shown that the muonic component still has to be scaled up to match observed data, while no rescaling of the EM component and the FD energy is required~\cite{Aab:2016hkv}.
The measurements with the FD also show that both the position of the shower maximum in the atmosphere ($\xmax$) and the entire shape of the EM shower are well described by the simulations~\cite{Aab:2014kda,Aab:2018jpg}.
At lower energies, down to $\sim 10^{17.3}\,$eV, in a measurement using the subarray of buried scintillators of the Pierre Auger Observatory, a direct count of the muons independent of EM contamination was obtained, which also shows that simulations produce too few muons~\cite{Aab:2020frk}.
There is much evidence that all the simulations underpredict the average number of muons in the showers: a comprehensive study of muon number measurements made with different experiments has shown that the muon deficit in simulations starts around $\sim 10^{16}\,$eV and steadily increases with energy.
Depending on model and experiment, the deficit at $\sim 10^{20}\,$eV ranges between tens of percent up to a factor of 2~\cite{whispICRC2019}.

The increased statistics obtained at the Pierre Auger Observatory allow us to now take a further step and explore fluctuations in the number of muons between showers, hereinafter referred to as {\it physical fluctuations}.
The ratio of the physical fluctuations to the average number of muons (relative fluctuations) has been shown to be mostly dominated by the first interaction, rather than the lower energy interactions deeper in the shower development~\cite{Fukui:1960aa,Cazon:2018gww}.
Here, we exploit the sensitivity of fluctuations to the first interaction to explore hadronic interactions well above the energies achievable in accelerator experiments.

\section{Methodology}
\label{sec:measurement}
Our analysis here is based on the set of inclined air showers ($\minZenith{}<\theta<\maxZenith{}$) that are reconstructed both with the SD and FD between \dataStart{} and \dataEnd{}.
For each event, we obtain independent measurements of the muon content (with the SD) and the calorimetric energy (with the FD). 
To ensure the showers can be reconstructed with small uncertainties, we select only events with at least four triggered stations in the SD array and we further require that all the stations surrounding the impact point of the shower on the ground are operational at the time of the event. 
Only events with good atmospheric conditions (few clouds and a low aerosol content) are accepted in order to guarantee a good energy reconstruction with the FD. In addition, it is required that the entire shower profile and, in particular, $\xmax$ is within the field of view of our telescopes.
Since heavy primaries penetrate the atmosphere less than light ones, the acceptance with this selection would be mass dependent.
To avoid this bias, we constrain the field of view to the region where all values of $\xmax$ are accepted.
Further details are given in~\cite{Aab:2014pza,Hexpo_2011}.
These selection criteria result in a total number of events of \neventSel{}.

In addition, only events with energy larger than $4 \times 10^{18}\,$eV, which ensures full trigger efficiency of the SD~\cite{inclinedReco}, are used to extract the fluctuations (\nevent{}~events).

The number of muons is reconstructed by fitting a 2D model of the lateral profile of the muon density at the ground to the observed signals in the SD array.
The free parameters of the fit are the zenith and azimuth angles of the shower, the impact point of the shower on the ground (shower core position), and a normalization factor with respect to a reference muon density profile in simulated proton showers at $10^{19}\,$eV~\cite{inclinedReco}.
There exists a residual pure EM component in showers with low zenith angles and stations very close to the shower core position (at $400\,$m and $64^\circ$ it is $\sim6\%$), which has been subtracted using a parametrization~\cite{Valino:2009dv}.
The dimensionless normalization factor we obtain from the fit is then transformed to the dimensionless quantity $\rmu$, which is given by the integrated number of muons at the ground divided by a reference given by the average number of muons in simulated proton showers at $10^{19}\,$eV and the given zenith angle. 
At $10^{19}\,$eV and an inclination of $60^\circ$, $\rmu = 1$ corresponds to $2.148 \times 10^7$ muons.
For more details, see~\cite{Aab:2014pza}.
In the following, we refer to $\rmu$ as the number of muons for short.

The calorimetric energy of the air showers $E_{\rm cal}$ is reconstructed by integrating the longitudinal shower profiles observed with the FD~\cite{Abraham:2009pm,Aab:2018jpg}.
The total energy of the shower is then obtained by adding the average energy carried away by muons and neutrinos, the so-called invisible energy $E~=~E_{\rm cal}+E_{\rm inv}$.
At $10^{19}\,$eV, $E_{\rm inv}$ accounts for $\enMuFrac$ of the total energy in air showers~\cite{Aab:2019cwj,supplement,Barbosa:2003dc,Risse:2003fw,Pierog2005}.

\begin{figure}[h]
  \centering
  \includegraphics[width=\columnwidth]{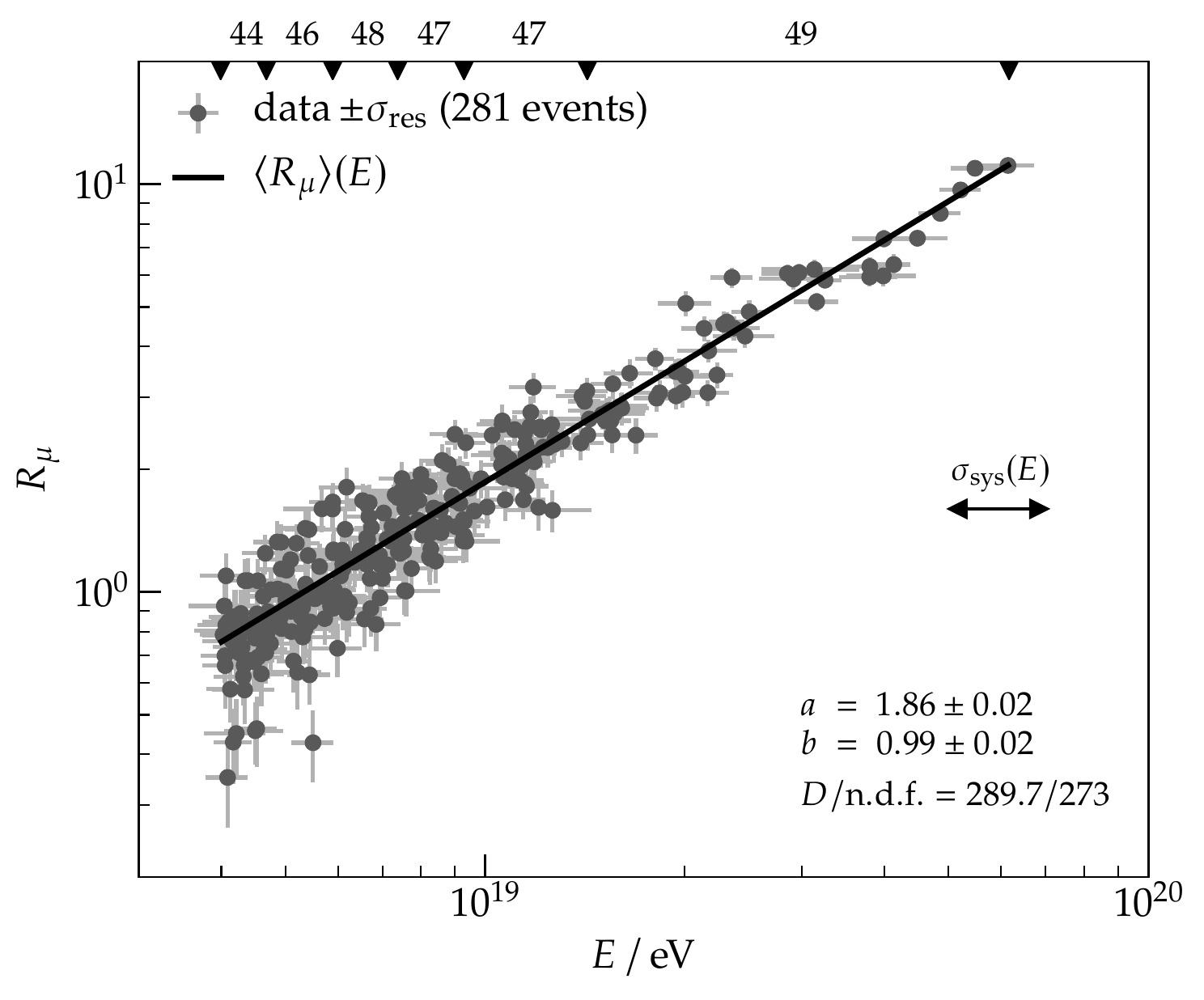}
  \caption{\label{fig:meas-calib}
    Number of muons as a function of the measured energy.
    The black line is the fitted $\mrmu=a\,[E/(10^{19}\,\mathrm{eV})]^b$.
    Markers on the top of the frame define the bins in which the fluctuations are evaluated.
    The numbers give the events in each bin.
    The effect of the uncertainty of the absolute energy scale is indicated by $\sigma_{\rm sys}(E)$.
  }
\end{figure}

In Fig.~\ref{fig:meas-calib} the muon number $\rmu$ is shown as a function of the measured energy. Markers on the top of the frame define the bins in energy for which we will extract the fluctuations, with the number of events in each bin shown above. The bins are chosen such that the number of events in each is similar.
Based on models of air shower development and given the gradual change of the composition in this energy range (single logarithmic dependence on energy)~\cite{Aab:2014kda,UngerKH,Matthews:2005sd,Engel:2011zzb}, the number of muons is related to the primary energy by a single power law

\begin{equation}
  \mrmu(E)=a [E/(10^{19}\,\mathrm{eV})]^b \ ,
  \label{eq:powerlaw}
\end{equation}
which can be fitted following a procedure described in the text below. The best-fit parameters are given at the beginning of the next section.
The scattering in the data has three sources: experimental uncertainties in the energy $\sre$ and in the muon number $\srmu$ from event reconstruction (both represented by the error bars), and the \textit{physical fluctuations} in the muon number denoted as $\sigma$.
Given Eq.~\eqref{eq:powerlaw}, the variance of the muon number is $\sigma^2 + b^2 \langle \sre \rangle^2 + \langle \srmu \rangle^2$.

In this Letter, we adopt a method based on maximizing the likelihood of a probability distribution function (PDF). The PDF incorporates the various contributions to the fluctuations, treating each energy bin independently while also accounting for the effect of the migration of events between bins~\cite{Aab:2014pza,Dembinski:2015wqa}.

The model assumes that measurements of $E$ and $\rmu$ follow Gaussian distributions centered at the true value, with widths given by the detector resolution $\sre$ and $\srmu$, which are the uncertainties obtained in each individual event reconstruction~\cite{inclinedReco,FDenergyICRC2019}.
Physical fluctuations are also assumed to follow a Gaussian distribution of width $\sigma$.
Simulations have shown this is an acceptable approximation given the event number in each bin.

The total PDF is obtained through the convolution of the detector response and the physical fluctuations with the probability distribution of the hybrid events measured at the Pierre Auger Observatory. The log-likelihood function is then given by

\begin{align}
   \ln \mathcal{L}(a,b,\hat{\sigma}_1,\ldots,\hat{\sigma}_6)~&=~\sum_i \ln \left [ \sum_{k=0}^{6} \, \int\limits_{E_{k-1}}^{E_k} \mathrm{d}E \, h(E) \, C(E) \, \right. \nonumber \\ 
   & \left. \times \exp{ \left ( -\frac{1}{2} \frac{(E_i -E)^2 }{s^2_{E}} \right ) }
   \right. \nonumber \\ 
   &  \left. \times \exp{ \left ( -\frac{1}{2 } \frac{(  R_{\mu,i}-\mrmu(E)\,)^2 }{s^2_{\mu} + (\hat{\sigma}_k \cdot \mrmu(E))^2 } \right ) } \right ] \ .
  \label{eq:likelihood}
\end{align}
The probability of hybrid events $h(E)$ (product of the energy spectrum of cosmic rays and the efficiency of detection) can be obtained from the data, as explained in~\cite{Dembinski:2015wqa} and~\cite{Aab:2020frk,SpectrumPRD}.
The rhs of Eq.~\eqref{eq:likelihood} depends on the parameters $a$ and $b$ via Eq.~\eqref{eq:powerlaw}. 
To obtain the energy dependence of the fluctuations, we parametrize $\sigma$ by six independent values such that $\sigma(E)~=~\hat{\sigma}_k \cdot \mrmu(E)$ where the constants $\hat{\sigma}_k$ are the relative fluctuations in the $k$th energy bin with limits $[E_{k-1}, E_k]$, where $k$ runs from one to six.
In Eq.~\eqref{eq:likelihood}, $k=0$ corresponds to the contributions from the interval $[0,E_{\rm thr}]$ where the SD is not fully efficient.
The fluctuations here are assumed to take the value of the first fitted bin $\hat{\sigma}_0 \equiv  \hat{\sigma}_1$.

The sum over the index $i$ in Eq.~\eqref{eq:likelihood} (the usual sum over the log-likelihoods of events) includes only events above the energy threshold of $4 \times 10^{18}\,$eV. The function $C(E)$ is the normalization factor from the double Gaussian.
The result of the fit 
for the parameters $a$ and $b$ are shown in Fig.~\ref{fig:meas-calib}.
The fluctuations are shown in Fig.~\ref{fig:result-sigma}.
The distribution of the number of muons and the PDF in the individual energy bins can be found in the Supplemental Material~\cite{supplement}.

The dominant systematic uncertainties of $\sigma$ come from the uncertainties in the resolutions $\sre$ and $\srmu$.
For $\srmu$ we estimate the uncertainty using simulations and data.
In simulations, the uncertainty was estimated by the spread in a sample of simulated showers, where each shower is reconstructed multiple times, each time changing only the impact point at the ground.
For data, we reconstruct the same event multiple times, leaving out the signals from one of the detector stations.
The average relative resolution $\langle \srmu/\rmu \rangle$ and its systematic uncertainty is thus \mRmuRes at $10^{19}\,$eV.

We verified the values of $\sre$ by studying the difference in the energy reconstruction of events measured independently by two or more FD stations. The width of the distribution of these energy differences is found to be compatible with $\sre$. We therefore take the statistical 1-$\sigma$ uncertainties of this cross check as a conservative upper limit of the systematic uncertainty of $\sre$~\footnote{The resolution systematics estimated in~\cite{FDenergyICRC2019} are smaller by about a factor three, but were derived for different quality cuts than the ones applied here.}. The average relative energy resolution $\langle s_E/E\rangle$ is about \mEnRes at $10^{19}\,$eV.

We have further confirmed that there are no significant contributions to the fluctuations from differences between the individual FD stations, neither related to the longtime performance evolution of the SD and FD detectors.

Any residual electromagnetic component in the signal would affect the lower zenith angles more.
We therefore split the event sample at the median zenith angle ($66^\circ$) and compare the resulting fluctuations.
We find no significant difference between the more and the less inclined sample.

In another test, we do find a small modulation of $\mrmu$ with the azimuth angle ($<1\%$), which we correct for.
This modulation is related to the approximations used in the reconstruction, which deal with the azimuthal asymmetry of the muon densities at the ground due to the Earth's magnetic field~\cite{inclinedReco}.
Finally, we have run an end-to-end validation of the whole analysis method described in this Letter on samples of simulated proton, helium, oxygen and iron showers.

Because of the almost linear relation between $\rmu$ and $E$, the systematic uncertainty on $\sigma$ due to the uncertainty of the absolute energy scale of $14\,$\%~\cite{FDenergyICRC2019} practically cancels out in the relative fluctuations. The systematic uncertainty in the absolute scale of $\rmu$ of $11\,$\%~\cite{Aab:2014pza} drops out for the same reason. The systematic effects for the bin around $10^{19}\,$eV are summarized in Table~\ref{tab:systematics}. Over all energies, the systematic uncertainties are below $\maxSystUncertainty$.

\begin{table}
    \begin{ruledtabular}
      \caption{\label{tab:systematics} Contributions to the systematic uncertainty in the relative fluctuations around $10^{19}\,$eV ($10^{18.97}\,$eV to $10^{19.15}\,$eV). The central value is $\sigma / \mrmu~=~0.102 \pm 0.029~(\mathrm{stat.}) \pm 0.007~(\mathrm{syst.})$.
      }
        \begin{tabular}{ccr}
        
Source of uncertainty & & Uncertainty  \\ 
\hline

$E$ absolute scale & $\langle E \rangle $ & $<0.1$ \% \\
$E$ resolution & $s_E$ & 4.6 \% \\
$R_{\mu}$ absolute scale & $\mrmu$ & 0.5 \% \\
$R_{\mu}$ resolution & $s_{\mu}$ & 5.2 \% \\
$\rmu$ azimuthal modulation & $ \mrmu(\phi)$ & 0.5 \% \\
\hline
 Total systematics& & 7.0 \% \\

        \end{tabular}
    \end{ruledtabular}
\end{table}

\section{Results and discussion}
\label{sec:results}

\begin{figure}
  \centering
  \includegraphics[width=\columnwidth]{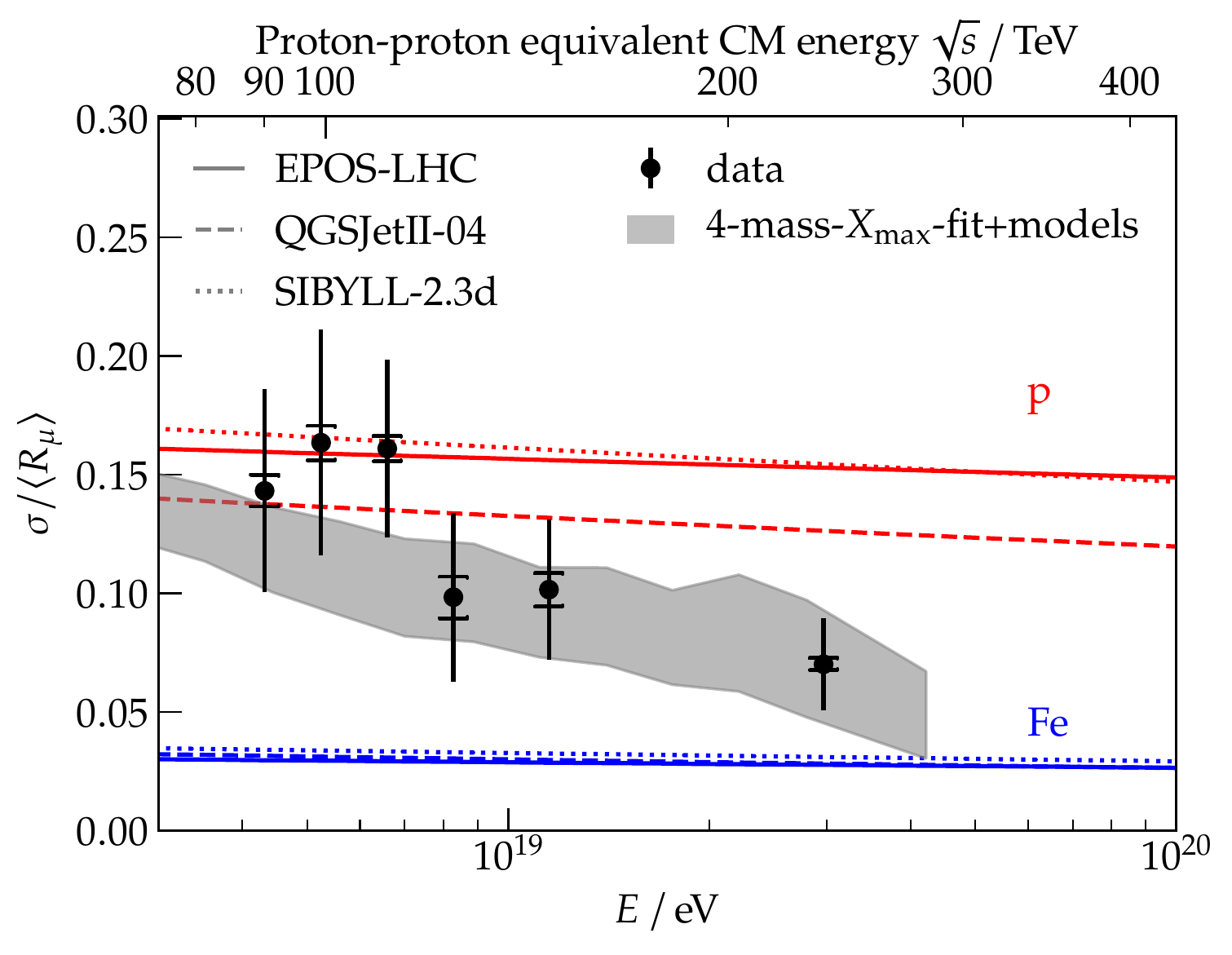}
  \caption{\label{fig:result-sigma} Measured relative fluctuations in the number of muons as a function of the energy and the predictions from three interaction models for proton (red) and iron (blue) showers. The gray band represents the expectations from the measured mass composition interpreted with the interaction models. The statistical uncertainty in the measurement is represented by the error bars. The total systematic uncertainty is indicated by the square brackets.}
\end{figure}

The best-fit value for the average relative number of muons at $10^{19}\,$eV (parameter $a$) is $\mrmu(10^{19}\,\si{eV}) = \mesA$.
For the slope (parameter $b$) we find $\de \mlnrmu / \de \ln E = \mesB$. These values are consistent with the values previously reported~\cite{Aab:2014pza,supplement}.

The measured relative fluctuations as a function of the energy are shown in Fig.~\ref{fig:result-sigma}.
We note that the measurement falls within the range that is expected from current hadronic interaction models for pure proton and pure iron primaries~\cite{Pierog:2009zt,Pierog:2013ria,Ostapchenko:2010vb,Ahn:2009wx,PhysRevD.102.063002,Bergmann:2006yz,Pierog:2004re,fluka,fluka2}.
To estimate the effect of a mixed composition, we take the fractions of the four mass components (proton, helium, nitrogen and iron) derived from the $\xmax$ measurements~\cite{Aab:2014kda,Bellido:2017cgf,sibyllFractionComment} and, using the simulations of the pure primaries, calculate the corresponding fluctuations in the number of muons.
The gray band in Fig.~\ref{fig:result-sigma} encompasses the predicted $\sigma/\mrmu$ of the three interaction models \qgs, \epos, and \sib~2.3d given the inferred composition mix for each~\cite{supplement}.

\begin{figure}
  \centering
  \includegraphics[width=\columnwidth]{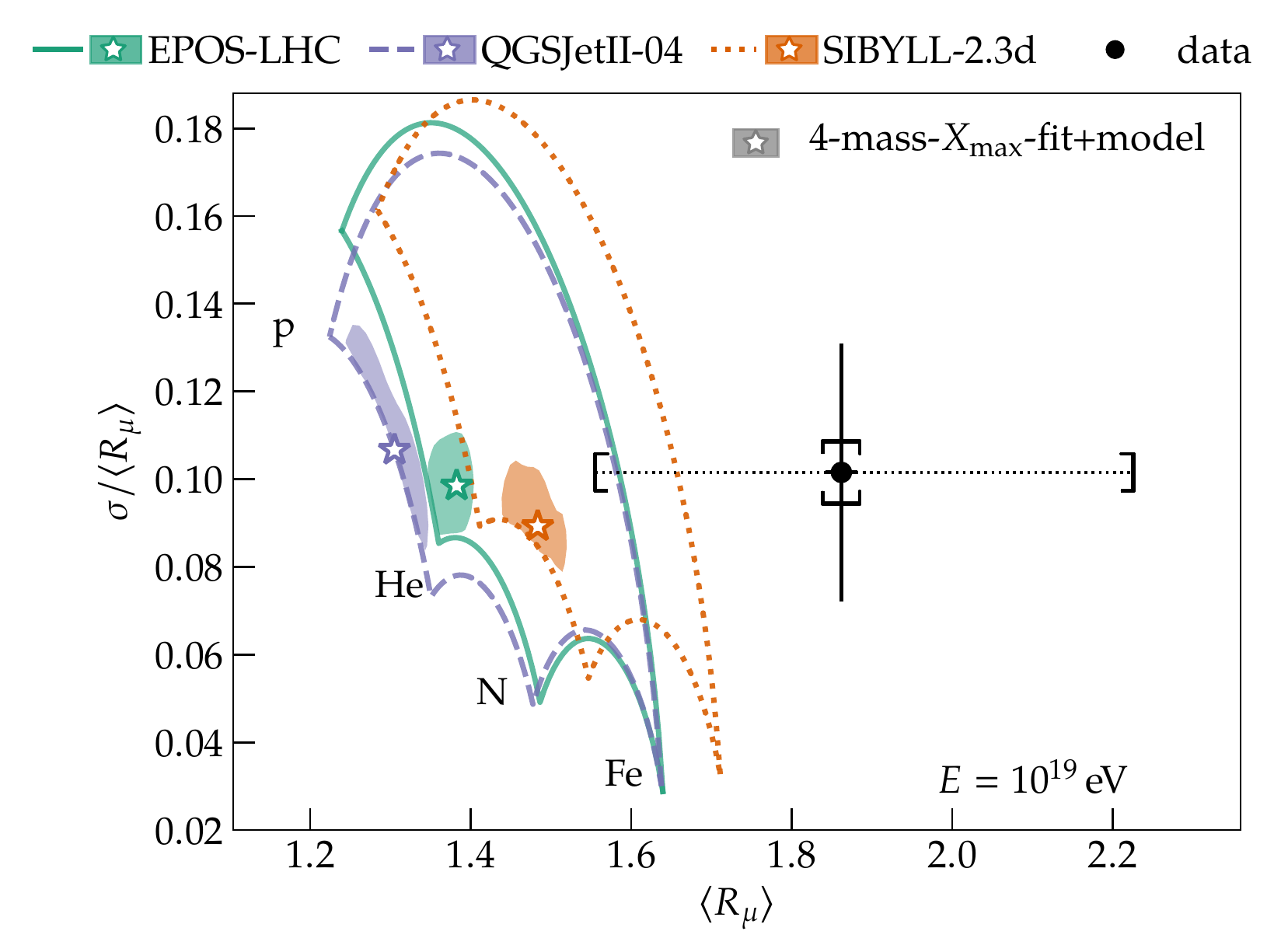}
  \caption{\label{fig:result-umbrella}
    Data (black, with error bars) compared to models for the fluctuations and the average number of muons for showers with a primary energy of $10^{19}\,$eV.
    Fluctuations are evaluated in the energy range from $10^{18.97}\,$eV and $10^{19.15}\,$eV.
    The statistical uncertainty is represented by the error bars.
    The total systematic uncertainty is indicated by the square brackets.
    The expectation from the interaction models for any mixture of the four components $p$, He, N, Fe is illustrated by the colored contours.
    The values preferred by the mixture derived from the $\xmax$ measurements are indicated by the star symbols.
    The shaded areas show the regions allowed by the statistical and systematic uncertainties of the $\xmax$ measurement~\cite{contourComment}.}
\end{figure}

In Fig.~\ref{fig:result-umbrella}, the effects of different composition scenarios on both the fluctuations and the average number of muons can be shown by drawing, at a fixed primary energy of $10^{19}\,$eV, the relative fluctuations $\sigma/\langle R_\mu \rangle$ against the average number of muons $\langle R_\mu \rangle$. Given any one of the interaction models, any particular mixture of the four components p, He, N, and Fe falls somewhere within one of the areas enclosed by the corresponding colored lines. The points of pure composition in this contour are labeled accordingly. For each model, the expected values for $\sigma/\langle R_\mu \rangle$ and $\langle R_\mu \rangle$ given the composition mixture obtained from the $X_{\rm max}$ measurements~\cite{Aab:2014kda} is indicated within each contour by the correspondingly colored star marker. The shaded areas surrounding the star markers indicate the statistical and systematic uncertainties inherited from the $X_{\rm max}$ measurements~\cite{contourComment}. Finally our measurement with statistical and systematic uncertainty is shown by the black marker.

Within the uncertainty, none of the predictions from the interaction models and the $X_{\rm max}$ composition (star markers) are consistent with our measurement. The predictions from the interaction models \qgs, \epos, and \sib~2.3d can be reconciled with our measurement by an increase in the average number of muons of \rescaleQGS{}, \rescaleEpos{}, and \rescaleSib{}, respectively.
For the fluctuations, no rescaling is necessary for any model.


Taken together, the average value and fluctuations of the muon flux constrain the way hadronic interaction models should be changed to agree with air shower data. To see this we briefly discuss the origin of the fluctuations.

The average number of muons in a proton shower of energy $E$ has been shown in simulations to scale as 
$\langle N^{*}_\mu \rangle= C E^\beta$ where $\beta \simeq 0.9$ ~\cite{Fukui:1960aa,Engel:2011zzb,Matthews:2005sd,Cazon:2018gww}. 

If we assume all the secondaries from the first interaction produce muons following the same relation as given for protons above, we obtain the number of muons in the shower as 
\begin{equation}
  N_\mu = \sum_{j=1}^{m} C~E_j^\beta =
 \langle N^{*}_\mu \rangle
 \sum_{j=1}^{m} x_j^\beta = \langle N_\mu^{*}  \rangle ~ \alpha_1 \ ,
  \label{eq:alpha1}
\end{equation}

where index $j$ runs over $m$ secondary particles which reinteract hadronically and $x_j=E_j/E$ is the fraction of energy fed to the hadronic shower by each \footnote{The energy fed to the electromagnetic shower in the first interaction is $E-\sum_{j=1}^m E_j$, and it rapidly decreases for next generations~\cite{Cazon:2019mtd}}. In this expression, the fluctuations in $N_\mu$ are induced by $\alpha_1$ in the first generation, which fluctuates because the multiplicity $m$ and the energies $x_j$ of the secondaries fluctuate~\cite{Cazon:2018gww}.

We can continue this reasoning for the subsequent generations to obtain 

\begin{equation}
\frac{N_\mu}{\langle N^{*}_\mu \rangle} = \alpha_1 \cdot \alpha_2 \cdots  \alpha_i \cdots \alpha_n \ ,
\label{eq:nmu-model}
\end{equation}

here the subindex $i$ runs over $n$ generations, until the cascade stops.
We note that, for the calculation of $\alpha_2$, in the second generation, there are $m$ particles contributing. 
Assuming the distributions of the $\alpha$'s for each one are similar, when adding up the muons produced by each, the fluctuations produced by one are statistically likely to be compensated by another. 
In other words, the $\alpha_2$ distribution is narrower by a factor $\sim 1/\sqrt{m}$. 
The deeper the generation, the sharper the corresponding $\alpha_i$ is expected to be. 
As a result, the dominant part of the fluctuations comes from the first interaction.
This has also been observed with simulations.
The model can be generalized for primary nuclei with mass $A$ using the superposition model and fixing the number of participants to $A$ protons, which reduces the different contributions to the fluctuations by a factor $\sim 1/\sqrt{A}$.

There are two options to increase the average number of muons in air showers.
One is to increase $\alpha$ in a specific generation, notably the first where the energy is the highest and exotic phenomena could conceivably play a role, i.e.\ $\alpha_1\to \alpha_1+\delta \alpha_1$. 
Note that, if only the first generation is modified (implying some sort of threshold effect for new physics), the increase in $\nmu$ is linear with the modification.
There are several examples in the literature where this approach has been used assuming different mechanisms~\cite{Aloisio:2014dua,AlvarezMuniz:2012dd,Anchordoqui:2016oxy,Farrar:2019cid,Farrar:2013sfa}.
For the fluctuations, the change depends on the model.
Alternatively, the number of muons can be increased by introducing small deviations in the hadronic energy fraction $\delta \alpha$ in all generations.
Accumulated along a number $n$ of generations, these small deviations build up as $\nmu \propto (\alpha+\delta \alpha)^n$.
For instance, a 5\% deviation per generation converts into $\sim 30\%$ deviation after six generations~\footnote{The number of generations is difficult to define as it depends on the details of the measurement and the showers, like the zenith angle, the muon energy threshold, distance from shower axis. Six is a reasonable minimum of generations before the muons reach the Auger detectors~\cite{Matthews:2005sd,AlvarezMuniz:2002ne}.}.
On the other hand, a change of 5\% in the fluctuations of $\alpha$ is not amplified in the muon fluctuations because of the suppression in later generations.
This approach characterizes the increase in the number of muons in the current hadronic interaction models with regard to previous models~\cite{Grieder:1973x1,Pierog:2006qv,Ostapchenko:2013pia,Drescher:2007hc,PhysRevD.102.063002}.
It is also compatible with the increase of the discrepancy in the average number of muons across a wide range of energies reported in~\cite{whispICRC2019}.

The present analysis finding that fluctuations are consistent with model predictions means that the increase in muon number may be a small effect accumulating over many generations or a very particular modification of the first interaction that changes $\nmu$ without changing the fluctuations~\cite{supplement}.

\section{Summary}
\label{sec:conclusion}
We have presented for the first time a measurement of the fluctuations in the number of muons in inclined air showers, as a function of the UHECR primary energy.
Within the current uncertainties, the relative fluctuations show no discrepancy with respect to the expectation from current high-energy hadronic interaction models and the composition taken from $\xmax$ measurements.

This agreement between models and data for the fluctuations, combined with the significant deficit in the predicted total number of muons, points to the origin of the models’ muon deficit being a small deficit at every stage of the shower that accumulates along the shower development, rather than a discrepancy in the first interaction.
Adjustments to models to address the current muon deficit must therefore not alter the predicted relative fluctuations.

The Pierre Auger Observatory is currently undergoing an upgrade that includes the deployment of scintillators on top of the SD stations~\cite{AugerPrime} to help disentangle the muonic and electromagnetic content of the showers, as well as an array of radio antennas~\cite{radioUpgrade}.
It has been shown that radio arrays can provide an estimate of the calorimetric energy~\cite{radiationEnergy}, and therefore, it will soon be possible to perform an analysis similar to the one presented here with much larger statistics using hybrid events measured by the high-duty-cycle radio and surface detector arrays~\cite{radioUpgrade}.

\begin{acknowledgments}
\input{acknowledgments}
\end{acknowledgments}

\input{letter.bbl.tex}

\clearpage
\onecolumngrid
\input{supplement.tex}

\end{document}

%% file: acknowledgments.tex

\section*{Acknowledgments}

\begin{sloppypar}
The successful installation, commissioning, and operation of the Pierre
Auger Observatory would not have been possible without the strong
commitment and effort from the technical and administrative staff in
Malarg\"ue. We are very grateful to the following agencies and
organizations for financial support:
\end{sloppypar}

\begin{sloppypar}
Argentina -- Comisi\'on Nacional de Energ\'\i{}a At\'omica; Agencia Nacional de
Promoci\'on Cient\'\i{}fica y Tecnol\'ogica (ANPCyT); Consejo Nacional de
Investigaciones Cient\'\i{}ficas y T\'ecnicas (CONICET); Gobierno de la
Provincia de Mendoza; Municipalidad de Malarg\"ue; NDM Holdings and Valle
Las Le\~nas; in gratitude for their continuing cooperation over land
access; Australia -- the Australian Research Council; Brazil -- Conselho
Nacional de Desenvolvimento Cient\'\i{}fico e Tecnol\'ogico (CNPq);
Financiadora de Estudos e Projetos (FINEP); Funda\c{c}\~ao de Amparo \`a
Pesquisa do Estado de Rio de Janeiro (FAPERJ); S\~ao Paulo Research
Foundation (FAPESP) Grants No.~2019/10151-2, No.~2010/07359-6 and
No.~1999/05404-3; Minist\'erio da Ci\^encia, Tecnologia, Inova\c{c}\~oes e
Comunica\c{c}\~oes (MCTIC); Czech Republic -- Grant No.~MSMT CR LTT18004,
LM2015038, LM2018102, CZ.02.1.01/0.0/0.0/16{\textunderscore}013/0001402,
CZ.02.1.01/0.0/0.0/18{\textunderscore}046/0016010 and
CZ.02.1.01/0.0/0.0/17{\textunderscore}049/0008422; France -- Centre de Calcul
IN2P3/CNRS; Centre National de la Recherche Scientifique (CNRS); Conseil
R\'egional Ile-de-France; D\'epartement Physique Nucl\'eaire et Corpusculaire
(PNC-IN2P3/CNRS); D\'epartement Sciences de l'Univers (SDU-INSU/CNRS);
Institut Lagrange de Paris (ILP) Grant No.~LABEX ANR-10-LABX-63 within
the Investissements d'Avenir Programme Grant No.~ANR-11-IDEX-0004-02;
Germany -- Bundesministerium f\"ur Bildung und Forschung (BMBF); Deutsche
Forschungsgemeinschaft (DFG); Finanzministerium Baden-W\"urttemberg;
Helmholtz Alliance for Astroparticle Physics (HAP);
Helmholtz-Gemeinschaft Deutscher Forschungszentren (HGF); Ministerium
f\"ur Innovation, Wissenschaft und Forschung des Landes
Nordrhein-Westfalen; Ministerium f\"ur Wissenschaft, Forschung und Kunst
des Landes Baden-W\"urttemberg; Italy -- Istituto Nazionale di Fisica
Nucleare (INFN); Istituto Nazionale di Astrofisica (INAF); Ministero
dell'Istruzione, dell'Universit\'a e della Ricerca (MIUR); CETEMPS Center
of Excellence; Ministero degli Affari Esteri (MAE); M\'exico -- Consejo
Nacional de Ciencia y Tecnolog\'\i{}a (CONACYT) No.~167733; Universidad
Nacional Aut\'onoma de M\'exico (UNAM); PAPIIT DGAPA-UNAM; The Netherlands
-- Ministry of Education, Culture and Science; Netherlands Organisation
for Scientific Research (NWO); Dutch national e-infrastructure with the
support of SURF Cooperative; Poland -Ministry of Science and Higher
Education, grant No.~DIR/WK/2018/11; National Science Centre, Grants
No.~2013/08/M/ST9/00322, No.~2016/23/B/ST9/01635 and No.~HARMONIA
5--2013/10/M/ST9/00062, UMO-2016/22/M/ST9/00198; Portugal -- Portuguese
national funds and FEDER funds within Programa Operacional Factores de
Competitividade through Funda\c{c}\~ao para a Ci\^encia e a Tecnologia
(COMPETE); Romania -- Romanian Ministry of Education and Research, the
Program Nucleu within MCI (PN19150201/16N/2019 and PN19060102) and
project PN-III-P1-1.2-PCCDI-2017-0839/19PCCDI/2018 within PNCDI III;
Slovenia -- Slovenian Research Agency, grants P1-0031, P1-0385, I0-0033,
N1-0111; Spain -- Ministerio de Econom\'\i{}a, Industria y Competitividad
(FPA2017-85114-P and FPA2017-85197-P), Xunta de Galicia (ED431C
2017/07), Junta de Andaluc\'\i{}a (SOMM17/6104/UGR), Feder Funds, RENATA Red
Nacional Tem\'atica de Astropart\'\i{}culas (FPA2015-68783-REDT) and Mar\'\i{}a de
Maeztu Unit of Excellence (MDM-2016-0692); USA -- Department of Energy,
Contracts No.~DE-AC02-07CH11359, No.~DE-FR02-04ER41300,
No.~DE-FG02-99ER41107 and No.~DE-SC0011689; National Science Foundation,
Grant No.~0450696; The Grainger Foundation; Marie Curie-IRSES/EPLANET;
European Particle Physics Latin American Network; and UNESCO.
\end{sloppypar}

%% file: letter.bbl.tex
%

%% file: supplement.tex
\section*{Supplemental material: Measurement of the fluctuations in the number of muons in extensive air showers with the Pierre Auger Observatory}

\subsection{Distribution of the number of muons and raw fluctuations}

\begin{figure*}[h]
  \centering
  \includegraphics[width=0.98\textwidth]{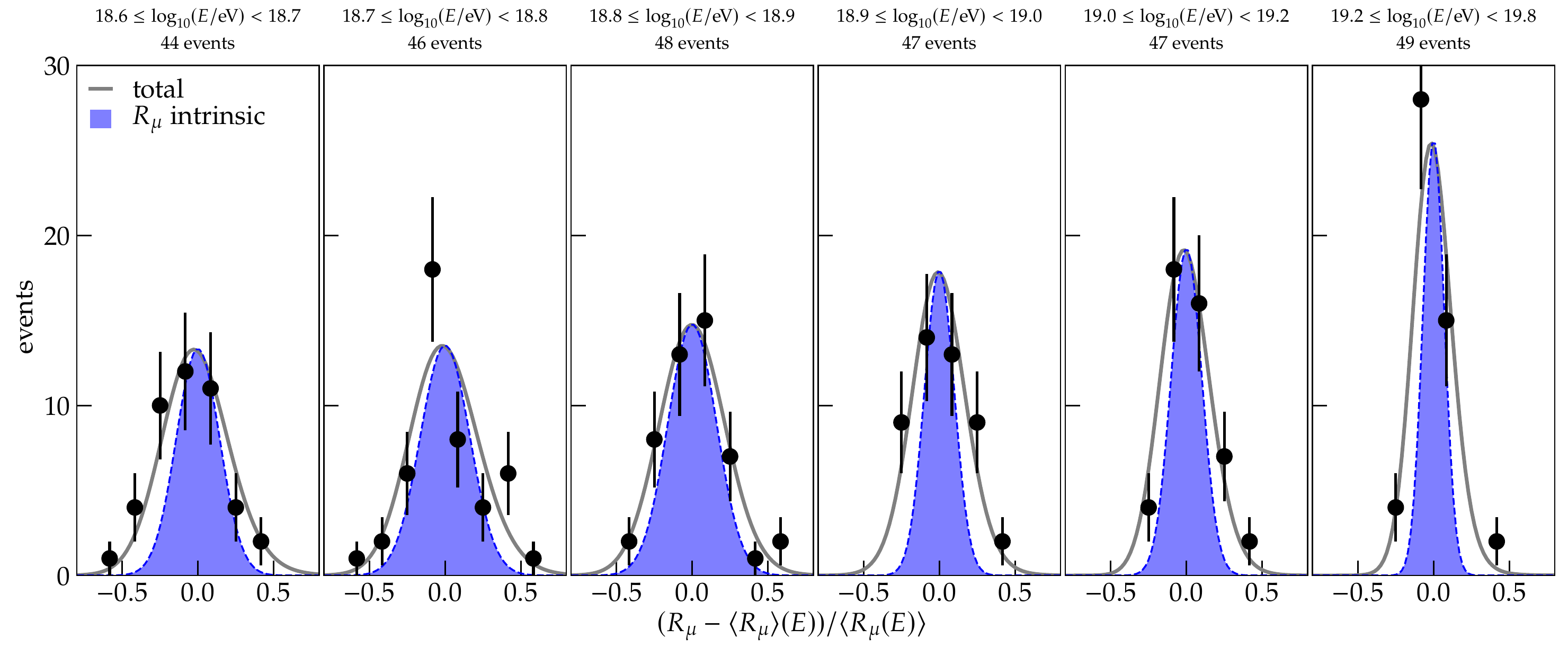}
  \caption{Distribution of the relative number of muons in six bins of energy from $10^{18.6}\,$eV to $10^{19.8}\,$eV. The model for the full distribution is shown in gray, the inferred intrinsic distribution of the number of muons is shown by the filled-in curve.}
  \label{fig:distributions}
\end{figure*}

The distribution of the relative number of muons $(\rmu-\mrmu)/\mrmu$ in the data in the six energy bins is shown in Fig.~\ref{fig:distributions}. The best-fit model for the data is shown in gray, the physical distribution is shown in blue. The data is well described by a Gaussian.

\begin{figure}[h]
  \centering
  \includegraphics[width=0.48\columnwidth]{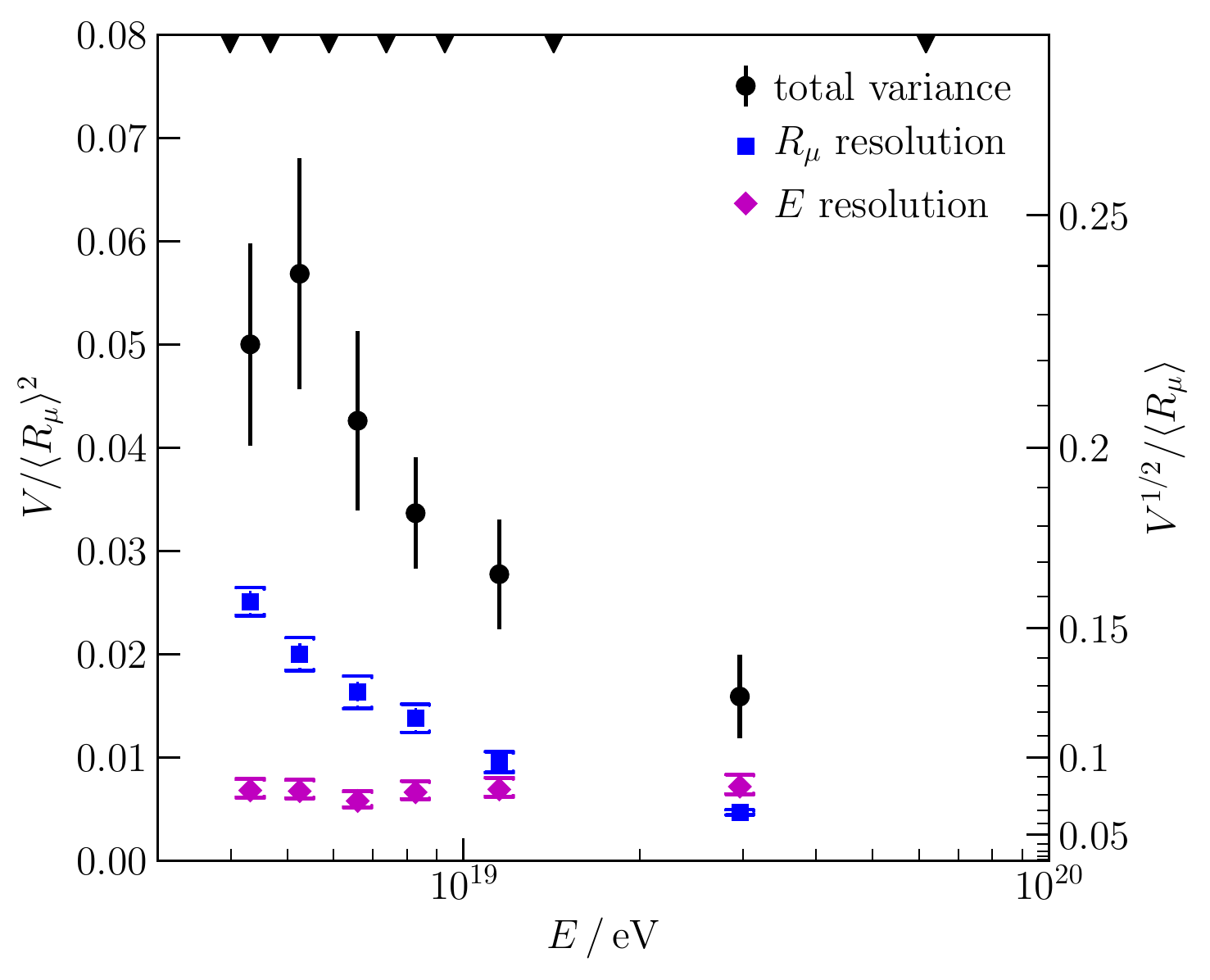}
  \caption{\label{fig:meas-variance} Black points show the total relative fluctuations in $\rmu$ as a function of the shower energy (left axis for the variance and right axis for the standard deviation). Blue and pink points show the average relative resolution in $\rmu$ ($\langle \srmu / \rmu \rangle$) and $E$ ($\langle \sre /E\rangle$) respectively. The error bars show the statistical uncertainty.}
\end{figure}

The relative variance in the data, $V/\mrmu^2$, and the average relative resolutions of the muon and energy measurements are shown in Fig.~\ref{fig:meas-variance}.

\newpage

\subsection{Detailed comparison between interaction models and measurement}
\begin{figure}
  \centering
  \includegraphics[width=0.48\columnwidth]{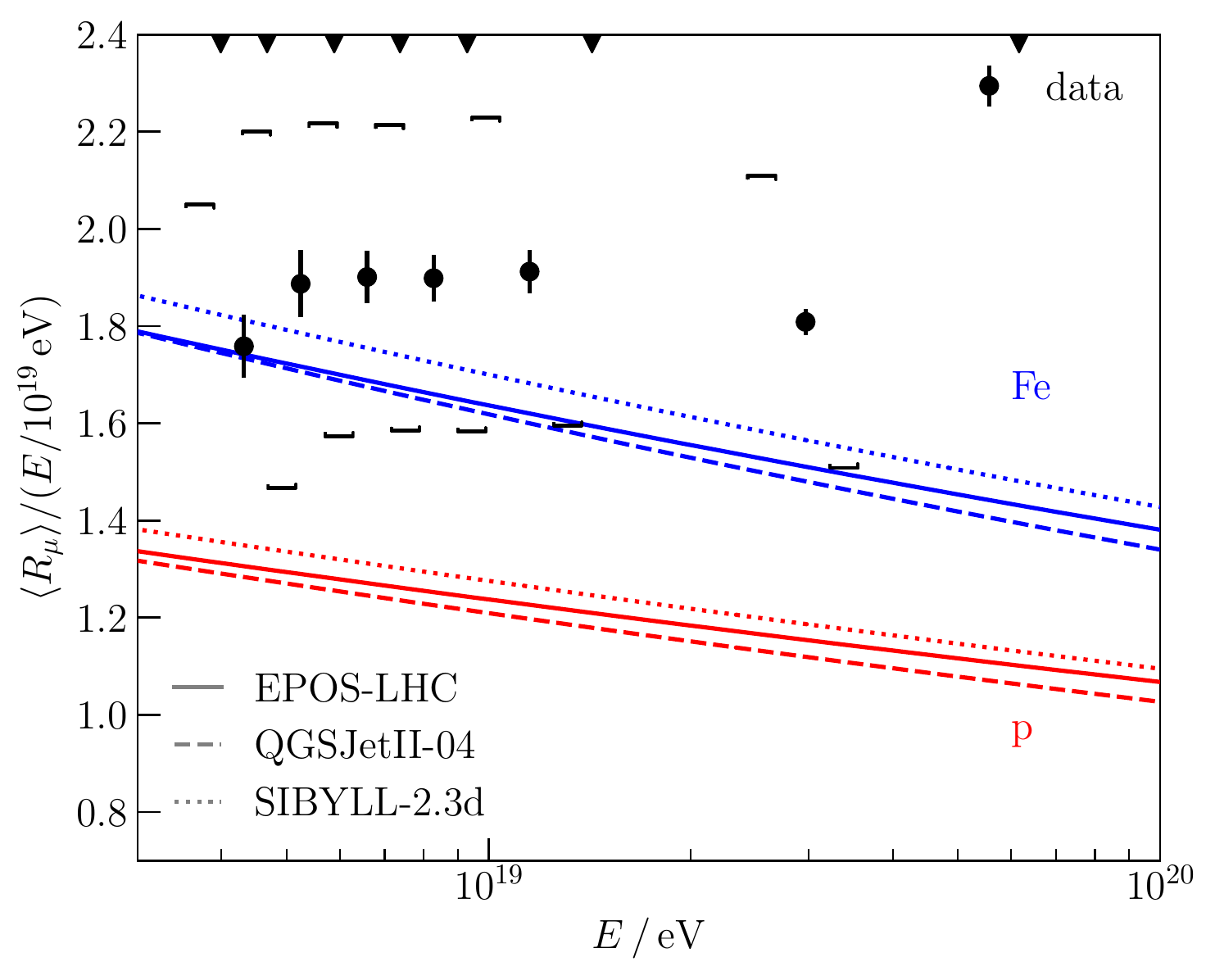}
  \caption{\label{fig:result-average} Measured average number of muons as a function of the energy and the predictions from three interaction models for proton (red) and iron (blue) showers.}
\end{figure}

In Fig.~\ref{fig:result-average} the average number of muons in each bin of energy is shown. 
The model predictions for proton and iron primaries are shown as well.

\begin{figure*}
  \centering
  \includegraphics[width=0.48\columnwidth]{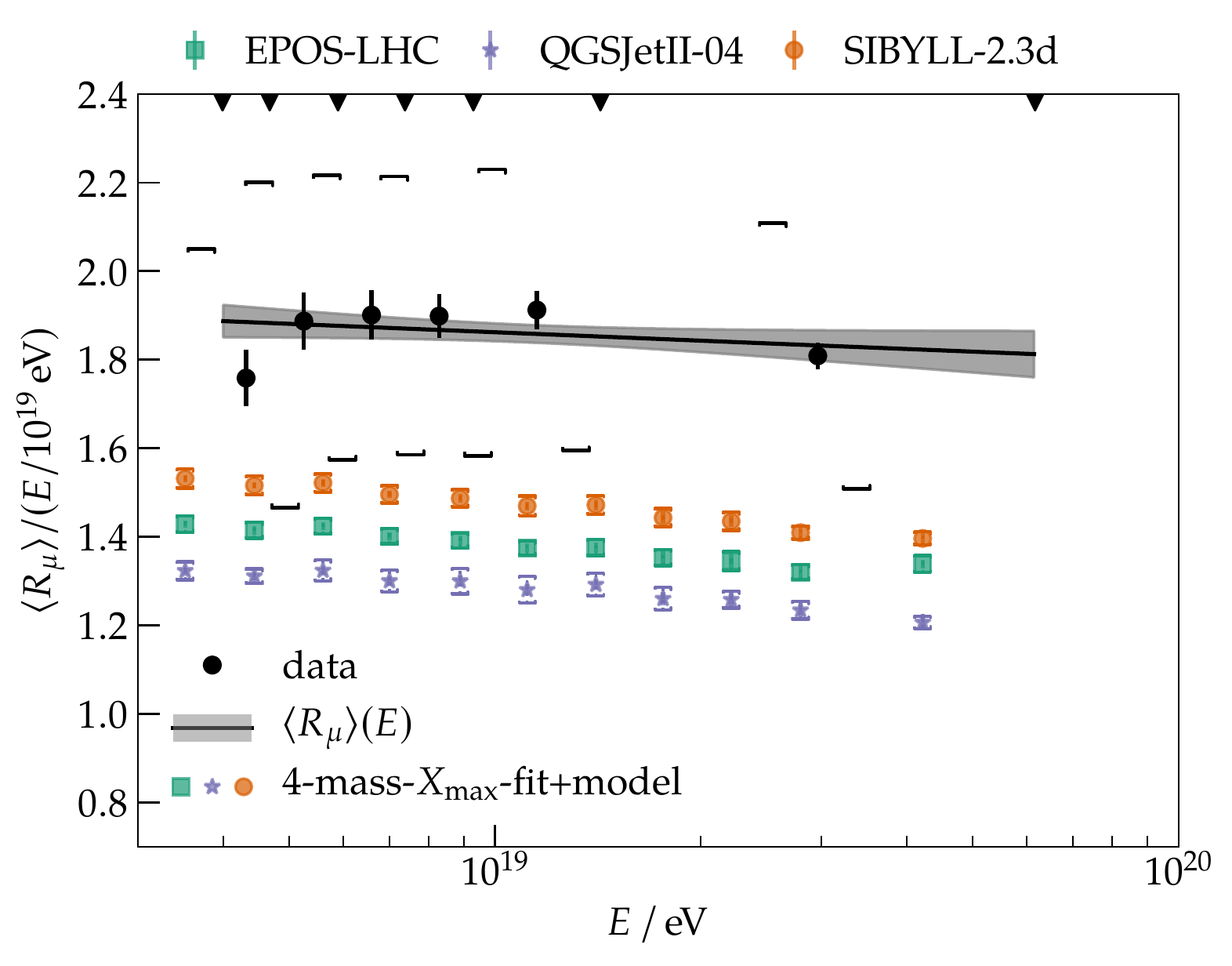}
  \hfill
  \includegraphics[width=0.48\columnwidth]{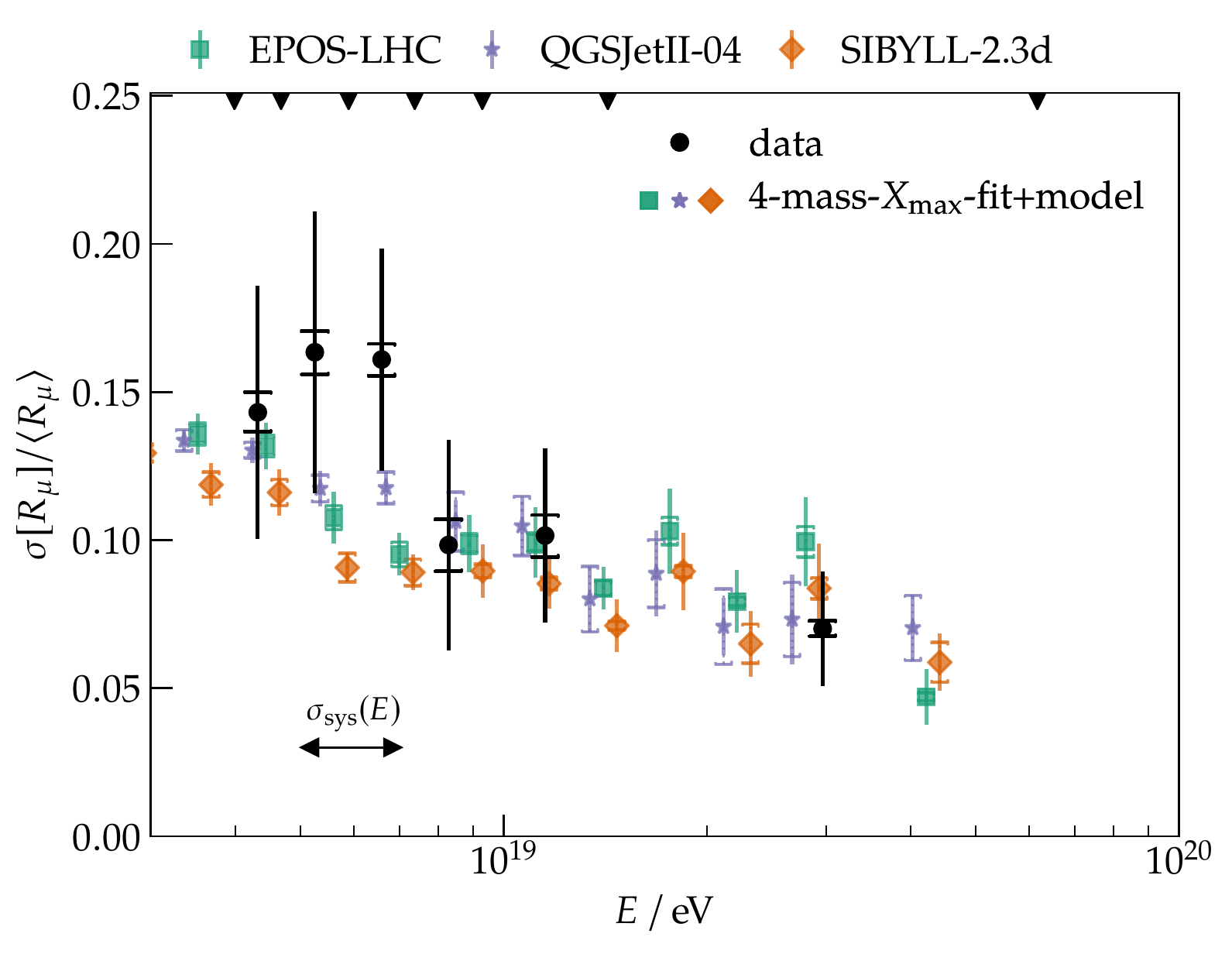}
  \caption{\label{fig:rmu-vs-xmax} Left panel: Average number of muons measured as a function of the energy together with the predictions from three interaction models given the composition measured with $\xmax$. The line is the best fit of the form $\mrmu[E]=a (E/(10^{19}\,\mathrm{eV}))^b$. Right panel: Relative fluctuations in the number of muons measured as a function of the energy together with the predictions from three interaction models given the composition measured with $\xmax$.}
\end{figure*}

\begin{figure*}
  \centering
  \includegraphics[width=0.98\textwidth]{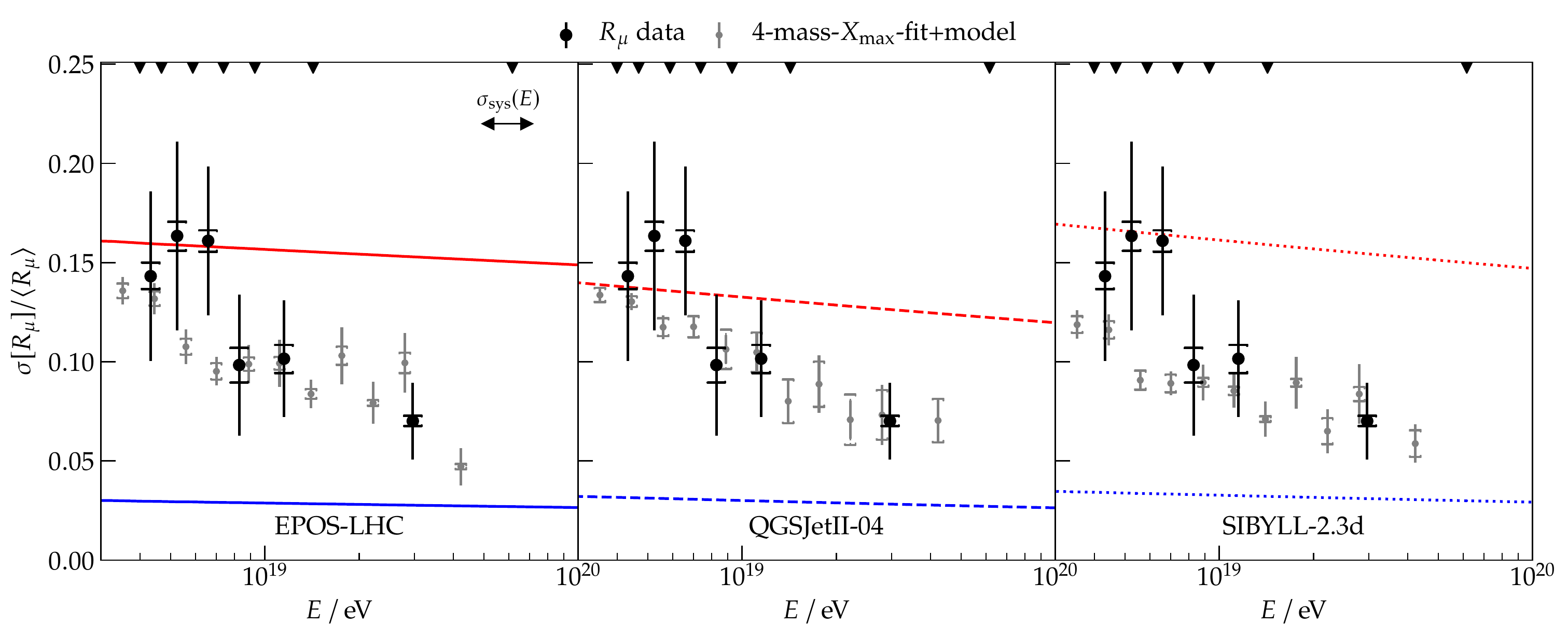}
  \caption{\label{fig:sig-rmu-xmax} Relative fluctuations in the number of muons measured as a function of the energy. The three panels show the predictions for the measured composition from \epos (left), \qgs (middle) and \sib~2.3d (right). The lines show the predictions for pure proton (red) and iron (blue).
    Fitting $\sigma(E)/\mrmu~=~p_0 + p_1 \, \log_{10}(E/10^{19}\mathrm{eV})$ to the measurement, yields $p_0~=~\sigFitA$ and the slope $p_1~=~\sigFitB$. The average slope predicted for pure proton (iron) primaries is $-0.01$ ($-0.003$). The values of $\chi^2/\mathrm{n.d.f.}$ between the trend expected from the measured composition and the measured fluctuations are $3.0/6$, $2.5/6$ and $4.3/6$ for \epos, \qgs and \sib~2.3d, respectively.
}
\end{figure*}

\begin{figure*}
  \centering
  \includegraphics[width=0.98\textwidth]{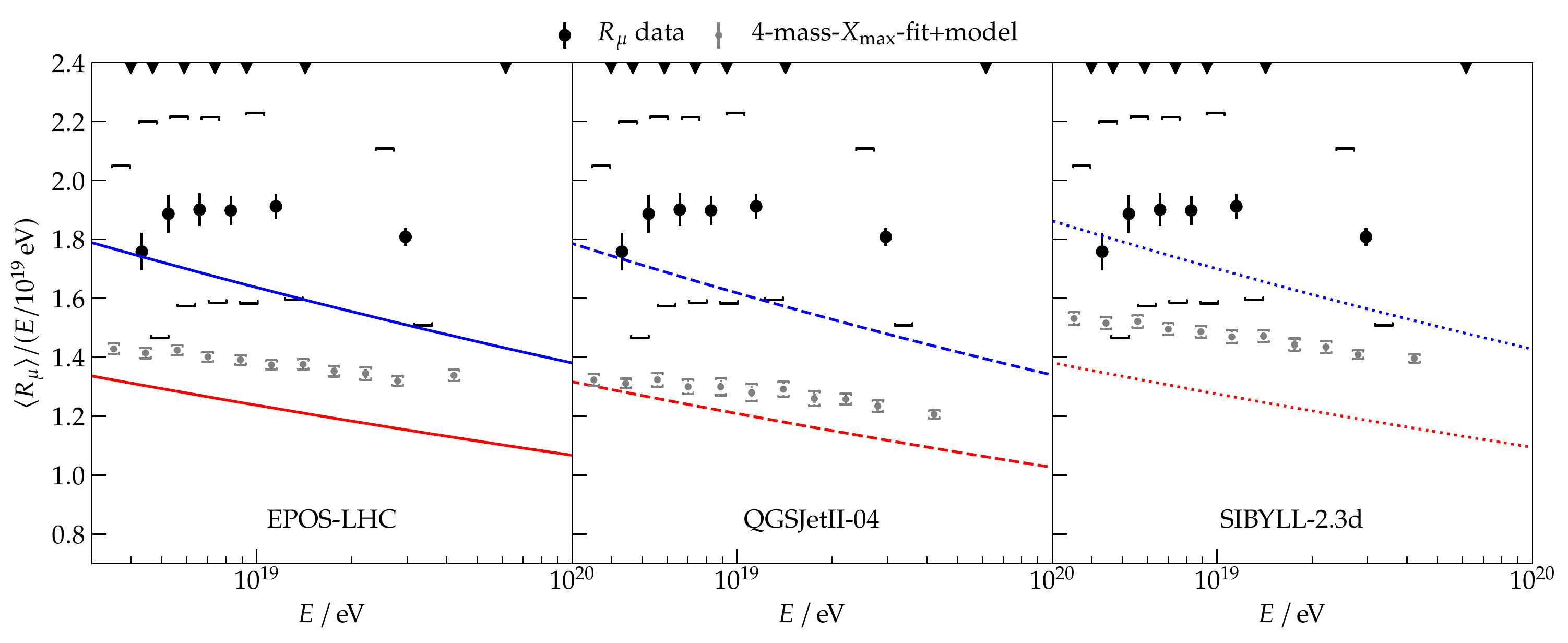}
  \caption{\label{fig:avg-rmu-xmax} Average number of muons measured as a function of the energy. The three panels show the predictions for the measured composition from \epos (left), \qgs (middle) and \sib~2.3d (right). The lines show the predictions for pure proton (red) and iron (blue). }
\end{figure*}

In Fig.~\ref{fig:rmu-vs-xmax} the measurement of the average number of muons (left panel) and the relative fluctuations (right panel) are shown as a function of the energy. The predictions from interaction models given the measured composition are shown for each model individually.
In Figs.~\ref{fig:sig-rmu-xmax} and~\ref{fig:avg-rmu-xmax} the measurement of the average number of muons and the relative fluctuations are compared with the predictions from the interaction models separately. All models, given the measured composition, reproduce the fluctuation measurement. In case of the average number of muons none of the models yields enough muons to describe the data.

\begin{figure}
  \centering
  \includegraphics[width=0.48\columnwidth]{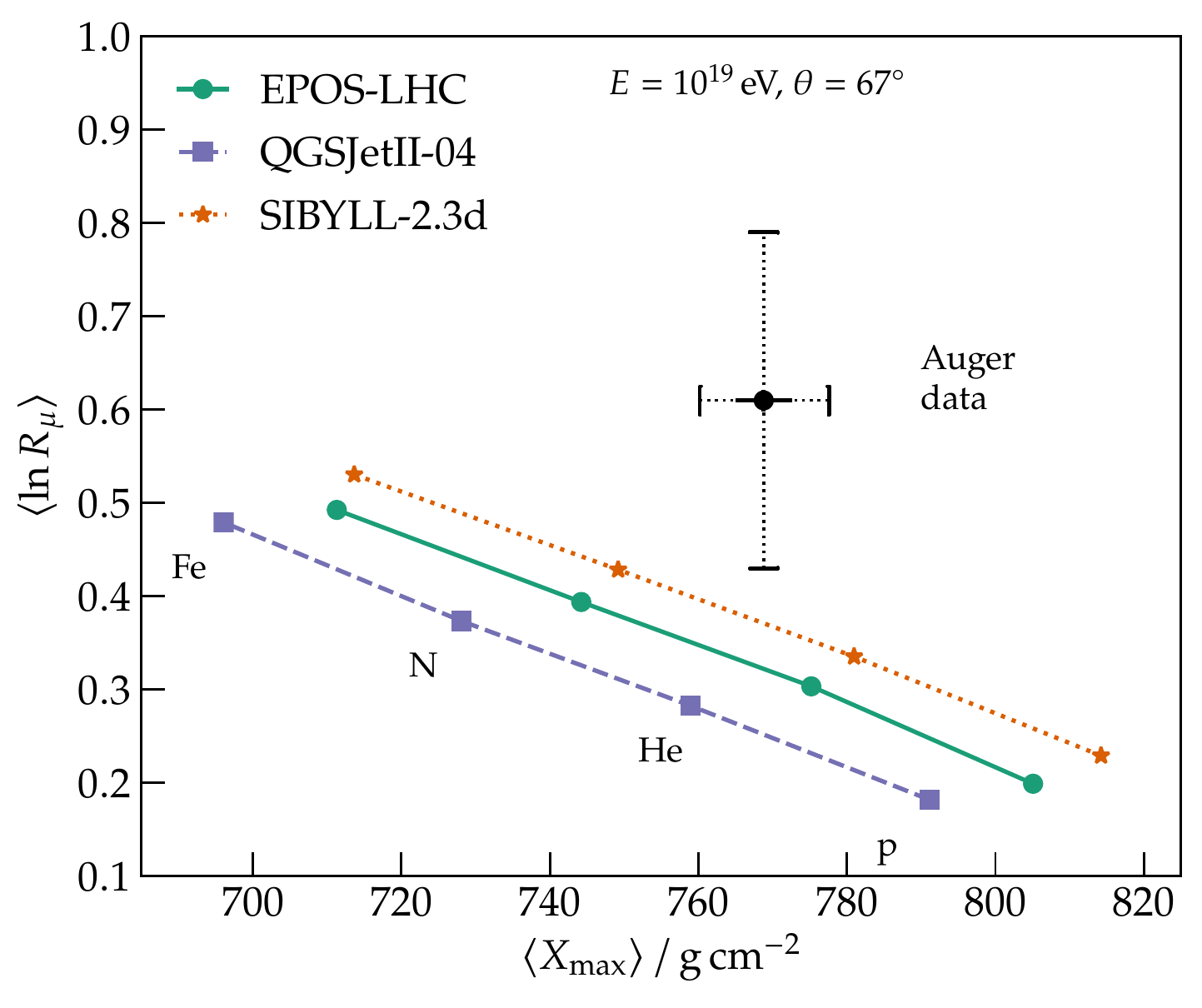}
  \caption{Average logarithmic muon content, $\mlnrmu$, as a function of the average shower depth, $\mxmax$.}
  \label{fig:avg-rmu-vs-avg-xmax}
\end{figure}

In Fig.~\ref{fig:avg-rmu-vs-avg-xmax} the measurement of $\mxmax$ and $\mlnrmu$ at $10^{19}\,$eV are compared. Both quantities scale linearly with $\langle \ln A \rangle$, meaning the predictions for different primary compositions fall on a line.

\subsection{Independence of the muon and energy measurements}

The direct contribution from muons to the calorimetric energy through the excitation of the molecules in the air is below $5$\% and thus the fluctuations introduced in $E_{\rm cal}$ by muons are negligible (see [18,19]).
For showers of a given total energy, due to the conservation of energy, $E_{\rm inv}$ and $E_{\rm cal}$ are anti-correlated on an event-by-event basis and $E_{\rm inv}$ depends on $\rmu$.
However, the fluctuations in $E_{\rm inv}$ due to the fluctuations in $\rmu$ are very small (around 1\% at $10^{19}\,$eV relative to $E$ (see [20])), such that in practice the determination of the two variables $E$ and $\rmu$ can be considered to be independent measurements. The value of $0.1$ we find for the relative fluctuations at $10^{19}\,$eV is consistent with this estimation of the fluctuations in the invisible energy.

\subsection{Number of muons and its fluctuations}

The average number of muons in a proton shower of energy $E$ has been shown in simulations to scale as 
$N^{*}_\mu(E)~=~C ~E^\beta$ where $\beta \simeq 0.9$ (see main text for references). 

If we assume all the secondaries from the first interaction produce muons following the same relation as given for protons above, we obtain the number of muons in the shower as 
\begin{equation}
  N_\mu(E) = \sum_{j=1}^{m} C~E_j^\beta =
 N^{*}_\mu(E)
 \sum_{j=1}^{m} x_j^\beta ~=~ N_\mu^{*}(E) ~ \alpha_1 \ ,
  \label{eq:alpha1}
\end{equation}
where index $j$ runs over $m$ secondary particles which reinteract hadronically and $x_j=E_j/E$ is the fraction of energy fed to the hadronic shower by each.
In this expression the fluctuations in $N_\mu$ are induced by $\alpha_1$ in the first generation which fluctuates because the multiplicity $m$ and the energies $x_j$ of the secondaries fluctuate.

Consider a ``toy`` interaction producing only pions, all with the same energy and only a fraction $f$ of them are charged and contribute to the hadron cascade. 
This model has no fluctuations and should by construction give $\alpha_1=1$, which follows from Eq.~\eqref{eq:alpha1} if we identify the average number of muons for proton showers with $N^{*}_{\mu}(E)$ which coincides with our definition.
This incidentally implies a condition for $\beta=\log (m)/ \log (m/f)$ which is the same as that obtained in [21,22] ($\beta \simeq 0.90$ for $f=2/3$ and $m\sim 50$).
In a more realistic scenario $\alpha_1$ fluctuates because the particles do not have the same energy and $f$ (the ratio of charged pions) and $m$ fluctuate.

%% file: letter.bbl
\begin{thebibliography}{57}
\bibitem [{sib()}]{sibyllFractionComment}%
  \BibitemOpen
  \href@noop {} {}\bibinfo {note} {{The mass fractions that are used for Sibyll
  here were derived for Sibyll~2.3. Given the small difference of
  $5\,$g$/$cm$^2$ in $\langle X_{\rm max} \rangle$ and $\sigma(X_{\rm max})$
  between Sibyll~2.3 and Sibyll~2.3d, the mass fractions for Sibyll~2.3d are
  expected to be very similar.}}\BibitemShut {Stop}%
\end{thebibliography}
